Luttinger liquid, singular interaction and quantum criticality in cuprate materials.

C. Di Castro and S. Caprara
Dipartimento di Fisica
Università di Roma "La Sapienza"
Piazzale Aldo Moro, 2
I-00185 Rome, Italy

Abstract

*With particular reference to the role of the renormalization group approach and Ward identities, we start by recalling some old features of the one-dimensional Luttinger liquid as the prototype of non-Fermi-liquid behavior. Its dimensional crossover to the Landau normal Fermi liquid implies that a non-Fermi liquid, as, e.g., the normal phase of the cuprate high temperature superconductors, can be maintained in d>1, only in the presence of a sufficiently singular effective interaction among the charge carriers. This is the case when, nearby an instability, the interaction is mediated by fluctuations. We are then led to introduce the specific case of superconductivity in cuprates as an example of avoided quantum criticality. We will disentangle the fluctuations which act as mediators of singular electron-electron interaction, enlightening the possible order competing with superconductivity and a mechanism for the non-Fermi-liquid behavior of the metallic phase.*
*This paper is not meant to be a comprehensive review. Many important contributions will not be considered. We will also avoid using extensive technicalities and making full calculations for which we refer to the original papers and to the many good available reviews. We will here only follow one line of reasoning which guided our research activity in this field.*

# 1. Introduction.

## 1.1 Criticality and Renormalization Group.

Up to the sixties of the last century the entire world of condensed matter or, more specifically, every *N*-body system in a stable phase was considered to be reducible to a collection of quasiparticles.

At the phenomenological level, a system of strongly interacting particles was considered at sufficiently low temperature as a gas of quasiparticles (quasiparticles of the normal Fermi liquid theory for metals or liquid $^3$He, phonons and rotons for superfluid Helium, gapped excitations for superconductors, spin waves for magnets…), plus at most a weak residual Hartree-type interaction.

Theoretically, it was privileged the development of techniques apt to obtain the simplified dynamics of quasiparticles for each system, starting from the microscopic description. The statistical part was then trivial, a gas of excitations with at most a normal distribution of fluctuations vanishing in the thermodynamic limit, according the $1/N^{1/2}$ law.

Out of this scheme, a puzzling behavior arises in the proximity of criticality (the first example of what is now called "complexity") where the collective phenomena do not appear as a simple superposition of single microscopic events and the laws of great numbers are modified (violation of the $1/N^{1/2}$ law). It is not true in this case that each sufficiently large portion of the system has an average behavior independent from the rest.

At zero temperature the competition between two ground states, rather than between two phases, and quantum fluctuations of the zero-point motion lead to quantum critical behavior near a Quantum Critical Point (QCP), which is obtained by tuning a parameter to balance the relative energies of the two states. Anomalous finite temperature behavior arises even though the transition is driven by non-thermal parameters, as applied and/or chemical pressure, doping, magnetic field, in fragile structures like liquid-solid helium, antiferromagnets and so on.

A common aspect of critical phenomena is the singular power-law behavior of the physical responses, like the spin susceptibility in ferromagnets near the Curie point, or the compressibility near the gas-liquid critical point. The power laws are characterized by sets of the so-called critical exponents, which are equal for apparently different transitions, like the para-ferromagnet or gas-liquid, provided the corresponding model systems, when expressed in terms of the order parameter $\varphi$, share the same symmetry. The order parameter characterizes the difference of the two phases (the spontaneous magnetization in the ferromagnet or the density difference near the gas-liquid critical point) and becomes the classical field of the field-theoretic models representing the system near criticality. A prototype of these models is the Landau-Wilson $\varphi^4$-model, generalization of the mean-field Landau theory of second-order phase transitions. This universal behavior observed near the critical point of phase transitions results from the fact that the correlation of the fluctuations of the order parameter extends to infinity and the system becomes scale invariant: Near a critical point, a change of the length scale leaves the physics unchanged, provided each parameter of the model is rescaled. The physical parameters $\mu_l$ are then characterized by their scaling dimension $x_l$ in terms of an inverse length.

Renormalization Group (RG) approach [1] implements these new universal simplifying concepts and asymptotically near criticality transforms into scaling transformation thus reproducing the phenomenology of criticality. The statistical aspects become prevalent with respect to the previous procedures in many body theory of approximately solving the dynamics, thus reducing each system to a gas of quasiparticles.

Two major RG transformations were introduced, each one implementing one aspect of universality.

While approaching the critical point, the details specifying each system do not matter [2]. The great (infinite) number of degrees of freedom involved as the correlation length $\xi$ of the order parameter goes to infinity, is well accounted for by a small set of parameters, which are considered to be relevant. Once the proper choice of the relevant parameters, e.g., the deviation from the critical temperature and the order parameter $\varphi$ (or its conjugate field $h$), is made, we can change the other parameters of the Hamiltonian, specifying the details of each system, and maintain the physics unchanged, provided we correspondingly renormalize the relevant parameters.

In 1969 the field theoretical RG approach was introduced in critical phenomena simultaneously in Russia [3] and in Rome [4]. The field theoretical RG implements this first form of universality and relates one model system to another by varying the coupling and suitably renormalizing the other variables and the correlation functions (vertices and propagators) to take care of the singularities in perturbation theory.

Since the correlation length $\xi$ goes to infinity, the degrees of freedom at short distance do not influence the critical behavior. The original system and the system expressed in terms of block-variables, related to larger and larger cells, should have the same critical behavior. In 1971 K. G. Wilson [5] implements this second idea of universality by introducing his new RG transformation with the elimination of degrees of freedom at large wavevectors, i.e. at short distances.

Both RG transformations aim to describe the system near the critical point in terms of few renormalized parameters, which correspond to a finite set of relevant directions with respect to a given fixed point. Relevant and renormalized parameters coincide asymptotically in the infrared region.

When the model parameters do not change anymore by iterating the transformation (fixed point), scaling and expressions of critical indices follow. Universality classes (yielding sets of critical indices common to different systems with the same symmetry) appear as domains of attraction of different fixed points.

The eigenvectors of the linearized RG transformation around each fixed point with negative scaling dimensions ($x_i<0$) define the tangent plane to the critical surface and correspond to the irrelevant variables. The eigenvectors with positive scaling dimensions ($x_i>0$) define the directions of escape from critical surface and are the relevant variables. The so-called marginal eigenvectors with scaling dimensions $x_i=0$ may occur. In this case, the critical indices or the scaling dimensions depend on the marginal parameters of the model, as in the Luttinger liquid (see Sec. 2.4).

## 1.2 Use of Renormalization Group for systems in stable phases and singular perturbation theory.

The most successful use of RG is indeed in critical phenomena with the asymptotic summation of infrared *singular* perturbative terms to give, for $d<4$, the singular power law behavior of physical response functions.

A different, more recent, application of RG deals with cases when perturbation theory is singular even in stable phases with *finite* physical response functions, e.g., for the two cases of interacting fermions at $d=1$ [6,7] and interacting bosons with Bose-Einstein condensation for $d\leq 3$ [8]. The case of interacting fermions at $d=1$ was the first case to put in crisis the picture of a collection of quasiparticles given by the Landau theory of normal Fermi liquid.

Both cases deal with stable liquid phases far from criticality and finite response functions, which therefore require exact cancellation of singularities to all orders in perturbation theory, instead of their summation into a power law singular behavior as in critical phenomena.

In critical phenomena, a general problem in the renormalization is to initialize the action to follow, in the parameter space of the Hamiltonian, a renormalized trajectory with a small number of flowing parameters compared to the enormous number of degrees of freedom strongly correlated within a coherence distance ($\xi \rightarrow \infty$). This subtle procedure of filtering out those variables appropriate to the description of critical systems and of their self organization cannot be carried out in general without a knowledge of the fundamental symmetry inherent to each specific problem, and in particular of the order parameter, to make the proper choice of the basic variables entering the field theoretic model, e.g. the Landau-Wilson model, on which the RG transformation acts.

Moreover, symmetry properties can be translated into Ward Identities (WI's), which establish relations among the various terms of the skeleton structure of the problem and their connection to physical quantities, thus in general simplifying the RG treatment.

In the case of singular perturbation theory in stable phases, additional symmetries and related WI's have to be present to implement the cancellation of singularities in physical responses, giving finite results despite a singular perturbation theory. In the cases of the Bose liquid (not treated here) [8] and of interacting fermions in $d=1$ (Luttinger model), the additional WI controlling this cancellation allows for a closure of the hierarchical equations of the response functions [6,10], thus providing the exact low-energy description of these systems. Exact solution of one-dimensional systems with forward scattering, the Luttinger liquid, was also previously obtained via bosonization in Refs. [11].

The crossover of the Luttinger liquid to the Fermi liquid in $d>1$ [7,12] and the non-Fermi-liquid behavior of an interacting electron system with singular forward interaction in $d>1$ [13,14] are also controlled by specific WI's and RG transformation. Extension of the bosonization technique to $d>1$ [15] was also used in this last case [16].

**1.3 Normal Fermi liquid and its breakdown in cuprates.**

In the absence of symmetry breaking, the description of the low energy behavior of metals in terms of a small set of parameters can be achieved via the RG scheme by an iterated elimination of high-energy degrees of freedom far from the Fermi surface (see, e.g., Ref. [17]).

The concept of a normal Fermi liquid, which successfully applies to liquid $^3$He and to ordinary metals in $d=3$, relies on the existence of a fixed point Hamiltonian with asymptotically free low-lying excitations, i.e., the quasiparticles, with discontinuous occupation number in momentum space $n_k$ at $T=0$ at the Fermi surface ($k=k_F$),

$$\left| n_{k_{F+}} - n_{k_{F-}} \right| = Z < 1,$$

where $k_{F\pm}$ indicates the limit $k \to k_F^{\pm}$ from above or below. This discontinuity still marks the Fermi surface in the interacting system and its *finite* reduction with respect to the Fermi gas ($Z=1$) is given by the finite "wavefunction renormalization" $Z$, renormalizing in this case the Fermi field operators. The presence of the discontinuity at the Fermi surface, together with the Pauli principle, compel the inverse quasiparticle lifetime to be $\tau^{-1} \approx \max(T^2, \varepsilon^2)$, where $\varepsilon$ is the deviation of the energy of the quasiparticle from the Fermi energy. The resistivity at finite temperature due to the electron-electron interaction is then proportional to $T^2$. The energy uncertainty $\hbar \tau^{-1}$ due to the finite lifetime of a quasiparticle near the Fermi surface is small compared to its energy $\varepsilon$ and the quasiparticle concept is well defined.

All the momentum transferring scattering processes become asymptotically ineffective and the expressions of the specific heat, the spin susceptibility and the compressibility or the Drude peak of a Fermi gas are still valid but for *finite* multiplicative renormalizations due to the residual Hartree type interaction among the quasiparticles.

The Fermi liquid breaks down if the quasiparticle spectral weight at the Fermi surface is suppressed by the presence of an interaction-induced anomalous scaling dimension $\eta$ in $Z \approx |k - k_F|^{\eta}$, where $k - k_F$ measures the deviation of the quasiparticle momentum from the Fermi momentum. Z vanishes at the Fermi surface, the low-lying single particle excitations are suppressed and the quasiparticle concept loses its validity. The low-energy behavior of the system is dominated by the charge and spin collective modes. These modes propagate with different velocities, leading to the so-called charge and spin separation, as opposed to the behavior of a normal Fermi liquid, where quasiparticle excitations simultaneously carry charge and spin. This non-Fermi-liquid-behavior is achieved in one-dimensional interacting electron system. The metallic phase is the so-called Luttinger liquid with a non–universal index $\eta$, finite charge and spin density responses in the long wavelength limit and the corresponding propagating low-lying collective modes.

The main motivation for discussing non-Fermi-liquid behavior in general comes from a number of experimental evidences in several systems. In particular the metallic phase of superconducting cuprates is not a Fermi liquid (see, e.g., Ref. [18]). These materials are insulating with antiferromagnetic long-range order when stoichiometric, and become strange metals and then superconducting upon chemical doping (see Sec. 4, Fig. 7). The cuprates are strongly anisotropic materials, and their structure consists of copper-oxygen planes intercalated with rare-earth slabs. The doped charge carriers are introduced in the copper-oxygen planes giving rise to quasi two-dimensional systems. When the chemical doping $x$ is such that the critical temperature $T_c$ for the onset of superconductivity reaches its maximum value (optimal doping), the in-plane resistivity is linear in temperature (rather than quadratic, as in Fermi liquids) over a wide temperature range. Below the optimal doping, in the so-called underdoped region, the anomalous metallic behavior is even stronger and a pseudogap opens below a doping dependent crossover temperature $T^*(x)$, affecting thermodynamic, spectroscopic, and transport properties. Furthermore, in contrast to the Fermi-liquid behavior, the inverse quasiparticle scattering time around optimal doping is linear in temperature or frequency, the energy uncertainty becomes of the same order of the energy, and the quasiparticle concept loses its validity. The appearance of a novel metallic phase, at least as relevant as the appearance of high temperature superconductivity itself, has opened a wide theoretical debate.

Among the various suggestions for a breakdown of the Fermi-liquid behavior [18,19], the possibility that a Luttinger-liquid-like phase could be extended to two dimensions has been considered [20]. We will indicate the constraints that have to be satisfied by the effective electron-electron interaction to extend the non-Fermi-liquid behavior to dimension higher than one. This led to consider as a further possibility for the occurrence of non-Fermi-liquid behavior that the system is close to an instability [18,21,22]. The nearby critical fluctuations couple to the charge carriers giving rise to an effective interaction among the fermionic quasiparticles singular enough to destroy the Fermi-liquid behavior.

Critical fluctuations can be due to a charge instability [incommensurate charge density wave [21] (CDW)] of the Fermi liquid coming from the markedly metallic overdoped region, to a magnetic instability [22] at low doping, or to a combination of the two, with a stripe like modulation of the charge and spin density [23]. We will deal with the quantum criticality related to the CDW as the onset of stripe phase in high-$T_c$ superconducting cuprates. Unveiling the critical modes that act as mediators of the effective interaction may also be a key to understand the elusive electronic order, which competes with superconductivity in the underdoped region [24].

**2. The Luttinger liquid and its crossover to the Fermi liquid.**

**2.1 The Tomonaga-Luttinger (TL) model.**

In one-dimensional systems the Fermi surface consists of two points $\alpha k_F$, where $\alpha=\pm$ refer to right- and left-moving fermions, respectively. The peculiar Fermi surface leads in general to consider few scattering processes that are incorporated in the continuous low-energy model known as the "g-ology model" [25]. This model includes the small momentum transfer scattering processes ($g_2$ and $g_4$), the back scattering ($g_1$) and Umklapp scattering ($g_3$), as depicted in Fig. 1. Many features of the g-ology model were already clear within perturbative renormalization calculations [25].

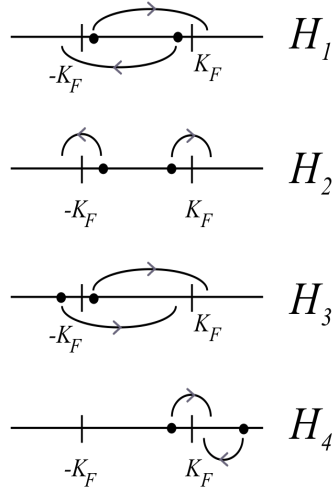

Fig. 1 - The relevant scattering processes considered within the g-ology model. The dots indicate the initial states of two scattering quasiparticles and the arrows point to their final states.

For the sake of simplicity, we will only refer to the Tomonaga-Luttinger (TL) [26] model with linear spectrum ($\varepsilon_k = v_F \mathbf{k}$), a bandwidth cut-off $\Lambda$ and the couplings $g_4$ and $g_2$ for the two forward scattering processes depicted in Fig. 1. The Hamiltonian $H = H_0 + H_2 + H_4$ ($H_0$ is the free particle Hamiltonian) is specified by

$$H_0 = \sum_{|\mathbf{k}|<\Lambda} \alpha v_F \mathbf{k} \, a^+_{\alpha,\sigma}(\mathbf{k}) a_{\alpha,\sigma}(\mathbf{k}),$$

$$H_2 = \frac{1}{V} \sum_{\mathbf{q}} \sum_{\sigma\sigma'} g_2^{\sigma\sigma'} \rho_{+,\sigma}(\mathbf{q}) \rho_{-,\sigma'}(-\mathbf{q}), \quad (1)$$

$$H_4 = \frac{1}{2V} \sum_{\mathbf{q}} \sum_{\sigma\sigma'} g_4^{\sigma\sigma'} \sum_{\alpha} \rho_{\alpha,\sigma}(\mathbf{q}) \rho_{\alpha,\sigma'}(-\mathbf{q}).$$

Henceforth, $v_F$ indicates the Fermi velocity. We adopt the boldface notation for the momentum $\mathbf{k}$, measured relative to $\pm k_F$, even in $d=1$, to distinguish from bivectors $k=(\mathbf{k},\varepsilon)$, which include the energy variable $\varepsilon$. The couplings may be spin dependent,

$$g_i^{\sigma\sigma'} = g_{i\|}\delta_{\sigma,\sigma'} + g_{i\perp}\delta_{\sigma,-\sigma'},$$

and

$$\rho_{\alpha,\sigma}(\mathbf{q}) = \sum_{\mathbf{k}} a^+_{\alpha,\sigma}(\mathbf{k}-\mathbf{q}) a_{\alpha,\sigma}(\mathbf{k}) \quad (2)$$

are the density operators for right ($\alpha=1$) and left ($\alpha=-1$) movers in terms of creation and destruction operators of spin ½ fermions with spin projection $\sigma$, $a^+_{\alpha,\sigma}(\mathbf{k})$ and $a_{\alpha,\sigma}(\mathbf{k})$. The discrete structure of the Fermi surface of one-dimensional fermions in the presence of forward scattering implies the separate charge and spin conservation in the low energy processes for particle near the left and right Fermi points respectively. The related symmetry [6,7,10] allows for the exact solution of the model, yielding the Luttinger liquid behavior. We will exploit the WI's following from the above symmetry to constrain the structure of the RG equations to have a line of fixed points with an anomalous dimension $\eta$ in $Z$ depending on the couplings $g_2$ and $g_4$. The same symmetry ensures the finiteness of the compressibility and the spin susceptibility, thus making possible a metallic phase with no quasiparticles as low-lying excitations.

We give now few technical definitions. The physical content of a system resides in the Green's functions defined as the ground-state expectation value (angular brackets) of the time-ordered operator products:

$$G^{(2n,l)}(\mathbf{k}_1,t_1,...;\mathbf{k}'_1,t'_1,...;\mathbf{k}''_1,t''_1,...) = (-i)^{n+l} \langle \mathcal{T}_t \, a_{\alpha_1\sigma_1}(\mathbf{k}_1,t_1)...a_{\alpha_n\sigma_n}(\mathbf{k}_n,t_n) \times a^+_{\alpha'_1\sigma'_1}(\mathbf{k}'_1,t'_1)...a^+_{\alpha'_n\sigma'_n}(\mathbf{k}'_n,t'_n)$$
$$\times \rho_{\alpha''_1\sigma''_1}(\mathbf{k}''_1 t''_1)...\rho_{\alpha''_l\sigma''_l}(\mathbf{k}''_l t''_l)\rangle \quad (3)$$

where $a^+$, $a$ and $\rho$ are the operators in the Heisenberg representation and $\mathcal{T}_t$ is the time-ordering operator, and analogously for the spin density operators. The single-particle Green's function (the propagator) is given by $G_{\alpha\sigma} \equiv G^{(2,0)}(\mathbf{k},t;\mathbf{k},0) = -i\langle \mathcal{T}_t \, a_{\alpha\sigma}(\mathbf{k},t) a^+_{\alpha\sigma}(\mathbf{k},0)\rangle$, where we used the fact that the expectation value vanishes for different $\alpha$, $\sigma$, and $\mathbf{k}$. The Green's functions in the frequency representation will be denoted by $G^{(2n,l)}(\{k\})$ and are obtained as a Fourier transform of Eq. (3) with respect to the time arguments. The single-particle Green's function for the free case is $G_0 = [\varepsilon - \alpha v_F \mathbf{k} + \alpha \, \text{sgn}(\mathbf{k}) i 0^+]^{-1}$, with $Z=1$. In the diagrammatic representation of perturbation theory, the single-particle Green's function is represented by a solid line and the interaction by a dashed line. In a given diagram, external lines correspond to ingoing and outgoing particles. The diagrams for $G^{(2n,l)}(\{k\})$ are classified in diagrams that can or cannot be divided into disjoint parts with cutting an internal line. The latter are called one-particle irreducible. By truncating the external lines of the one-particle irreducible diagrams for $G^{(2n,l)}(\{k\})$ one obtains the corresponding vertices $\Gamma^{(2n,l)}(\{k\})$. The four-point vertex $\Gamma^{(4)} \equiv \Gamma^{(4,0)}(\{k\})$ represents the fully dressed interaction. In order to extract the physical properties from the Green's functions, one should in principle solve an infinite set of hierarchical equations, since the equations of motion couple a given Green's function with higher order Green's functions and vertices. The additional WI's present in the TL model allow for a closure of this set and for its full solution.

Let us recall that for ordinary critical phenomena, the paradigmatic model is the Landau-Wilson $\varphi^4$-model:

$$H = \int (c|\nabla\varphi|^2 + r\varphi^2 + u\varphi^4) d^d x = \sum_\mathbf{k} [(ck^2+r)\varphi^2 + u\varphi^4].$$

The field $\varphi$ represents the order parameter of the broken-symmetry phase, $r \approx T-T_c$ is the deviation from the critical temperature, and $u$ is the coupling of the fields with respect to the free ($\varphi^2$) Gaussian case. Given $H$ to be dimensionless, when measured in units of the finite characteristic energy $k_B T_c$, it turns out that $r$ has bare canonical dimensions in inverse length equal to 2, the field $\varphi$ has bare dimensions $(d-2)/2$ and the coupling $u$ has bare dimension $4-d$. The perturbation theory with respect to the bare dimensionless coupling $u/|r|^{(4-d)/2}$ becomes meaningless at criticality ($r \to 0$) for $d<4$, and a renormalized theory is required. The infrared singularities for $d<4$ are manifest already in the first perturbative correction to the coupling $u$, i. e., to the four point vertex function depicted in Fig. 2(a). The loop correction consists of a momentum integration over two free Gaussian propagators (the two lines) $G_0=$

$(ck^2+r)^{-1}$ and a simple power counting gives four as upper critical dimensionality, where logarithmic singularities appear. Below d=4, stronger infrared singularities persist in perturbation theory and one needs the use of RG to take care of these singularities.

The singular behavior of perturbation theory in terms of the couplings $g_i$ for the TL model appears again by power counting at T=0. Indeed, a dimensional analysis of the model yields the canonical dimensions in inverse length: $[H]=1$, $[a(k)]=-1/2$ and $[g_i]=0$, i.e. $g_i$'s are marginal. In the one loop correction to the four-point vertex function [see Fig. 2(b)], one has $d$ for the momentum and $1$ for the frequency integration (at T=0, the time or its conjugate variable the frequency appear as additional variables), $-2$ for the two free propagators (the single particle Green function has in this case a linear dependence on momenta, rather then quadratic, and $d+1-2=d-1$). This shifts the logarithmic infrared singularities from $d=4$ (in the case of ordinary critical phenomena) to $d=1$ in the TL model.

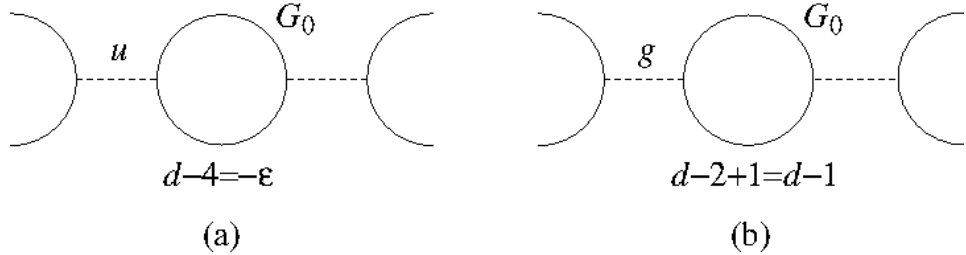

(a) (b)

Fig. 2 - (a) Power counting to determine the dimension of the effective coupling in critical phenomena. (b) Power counting to determine the dimension of the effective coupling in the TL model.

We will show that in this case one parameter Z, renormalizing the inverse single particle Green's function (the propagator), is enough to fully renormalize the TL model. In this case, Z vanishes at the Fermi points, driving the non-Fermi-liquid behavior.

## 2.2 The total charge and spin conservation and the related Ward identities.

The total charge (a=c) and spin (a=s) density operators $\rho^a(\mathbf{q})$ and the right and left charge and spin density operators $\rho^a_\alpha(\mathbf{q})$ are given by:

$$\rho^a(\mathbf{q}) = \sum_a \rho^a_\alpha(\mathbf{q}),$$

$$\rho^c_\alpha = \rho_{\alpha\uparrow}(\mathbf{q}) + \rho_{\alpha\downarrow}(\mathbf{q}), \quad (4)$$

$$\rho^s_\alpha = \rho_{\alpha\uparrow}(\mathbf{q}) - \rho_{\alpha\downarrow}(\mathbf{q}).$$

The right and left density operators $\rho^a_\alpha$ satisfy the commutation relation [6]:

$$\left[\rho^a_\alpha(\mathbf{q}), \rho^{a'}_{\alpha'}(\mathbf{q}')\right] = \frac{V}{\pi}\delta_{\alpha\alpha'}\delta_{\mathbf{q},-\mathbf{q}'}\alpha\mathbf{q}.$$

From the above commutation relation, the continuity equation related to the total charge (spin) conservation is obtained:

$$i\partial_t \rho^a(\mathbf{q},t) = [\rho^a(\mathbf{q},t), H] = \mathbf{q} v^a j^a(\mathbf{q},t), \qquad (5)$$

$$v^a = v_F + \frac{1}{\pi}(g_4^a - g_2^a),$$

where $j^a = \sum_\alpha \alpha \rho_\alpha^a$ is the current operator, which gives the physical current when multiplied by the interaction-dependent velocity $v^a$, and $g_i^{c,s} = \tfrac{1}{2}(g_{i\parallel} \pm g_{i\perp})$.

These conservations can be transformed into WI's by applying $i\partial_t$ to any correlation function and using the continuity equation. Without making explicit derivations, we summarize the main results [6,7]:

The total conservations connects vertices and correlations, e.g., the one particle Green's function $G$ is related with the charge (spin) density ($\mu$=0) and current ($\mu$=1) vertices $\Lambda_{\alpha,\sigma;\mu}^a(k,q)$:

$$\omega \Lambda_{\alpha,\sigma;0}^a(p,q) - v_F \mathbf{q} \Lambda_{\alpha,\sigma;1}^a(p,q) = \varepsilon_\sigma^a \left[ G_{\alpha,\sigma}^{-1}(p-q/2) - G_{\alpha,\sigma}^{-1}(p+q/2) \right], \qquad (6)$$

with $\varepsilon_\sigma^a = -1$ if $a=s$ and $\sigma=\downarrow$ and $\varepsilon_\sigma^a = +1$ otherwise. The truncated vertices $\Lambda_{\alpha,\sigma;\mu}^a(k,q)$ are obtained from the one-particle irreducible part of $G^{(2,1)}$ with $\rho_{\alpha\sigma}$ substituted by $\rho^a$ ($\mu$=0) and $j^a$ ($\mu$=1) in Eq. (3), respectively. These vertices are irreducible also with respect to cutting an interaction line. Notice that $v_F$, instead of $v^a$, appears in the WI, Eq. (6), for the irreducible vertices [6]. This WI is valid for all systems with total conservation. By itself is of course not enough to solve any model.

## 2.3 Left and right conservation laws and the additional Ward identities.

In addition to the usual total charge and spin conservation, the discrete structure of the Fermi systems in $d=1$ allows for more stringent conservation laws when large momentum scattering processes are absent, as in the TL model: charge (spin) density near the left and right Fermi points is conserved separately.

Let us introduce the so-called axial charge (spin) density as the difference operator $\tilde{\rho}^a = \rho_+^a - \rho_-^a = j^a$. These quantities also are now conserved and obey the continuity equation

$$i\partial_t \tilde{\rho}^a(\mathbf{q},t) = [\tilde{\rho}^a(\mathbf{q},t), H] = \mathbf{q} \tilde{v}^a \tilde{j}^a(\mathbf{q},t), \qquad (7)$$

with the axial current $\tilde{j}^a = \rho_+^a + \rho_-^a = \rho^a$ and the interaction-dependent velocity $\tilde{v}^a = v_F + (g_4^a + g_2^a)/\pi$.

One can define the correlation functions and the vertices for the axial density and current and, in complete analogy with the total charge ($a=c$) (spin, $a=s$) conservation, additional WI's are derived with $v^a$ substituted by $\tilde{v}^a$. The main consequences are:

1) Combining the WI's from global and axial conservation laws, the low frequency and momentum limits of the density-density and current-current correlation functions are now determined in terms of $v^a$ and $\tilde{v}^a$. The finite compressibility $\kappa$, spin susceptibility $\chi$ and the Drude weight $\mathrm{Re}\sigma(\omega)$ ($\sigma$ is the conductivity) are:

$$\kappa = 2/(\pi \tilde{v}^c), \quad \chi = 2/(\pi \tilde{v}^s), \quad \mathrm{Re}\sigma(\omega) = 2v^c \delta(\omega),$$

2) The continuity equations (5) and (7) for $\rho^a$ and $\tilde{\rho}^a$ are easily combined into a harmonic-oscillator equation for $\rho^a$:
$$i\partial_t \rho^a(\mathbf{q},t) + v^a \tilde{v}^a |\mathbf{q}|^2 \rho^a(\mathbf{q},t) = 0,$$
whence it follows that the undamped collective charge and spin modes move with different velocities $u^a = (v^a \tilde{v}^a)^{1/2}$ and the specific heat is linear in temperature:
$$C_V = \frac{\pi}{6} \frac{u^c + u^s}{u^c u^s} T.$$
The contribution to $C_V$ comes here from collective modes and not from quasiparticles, as in the Fermi liquid.

3) It is clear from the separate right and left conservation that the current vertex and the density vertex differ only from the sign of the particle velocity:
$$\Lambda^a_{\alpha\sigma;1}(p,q) = \alpha \Lambda^a_{\alpha\sigma;0}(p,q). \tag{8}$$
From Eq. (6) we can now eliminate the current vertex and obtain
$$(\omega - \alpha v_F \mathbf{q}) \Lambda^a_{\alpha\sigma;0}(p,q) = \varepsilon^a_\sigma \left[ G^{-1}_{\alpha\sigma}(p - q/2) - G^{-1}_{\alpha\sigma}(p + q/2) \right]. \tag{9}$$
This WI can be formally solved to express the density vertex, shortly named $\Lambda_0$ from now on, in terms of $G$ only. This is the key point to produce a closed solution for the single particle Green's function $G$, see section 2.5. We henceforth drop the index $\alpha$ to simplify the notation, making the dependence on $\alpha$ of the various quantities explicit when present.

## 2.4 The Luttinger liquid line of fixed points.

The only dramatic change introduced by the Luttinger liquid solution is in the single particle Green's function $G$, which maintains all the singular aspects of the problem. We have to show that no other renormalization is required besides the renormalization $Z$ for $G$. The bare couplings $g_i$ are dimensionless. From the perturbative corrections to the four-point vertex $\Gamma^{(4)}$, one sees that, at each order, a vertex and an integration over two propagators $G$ are added. This is also the skeleton structure of the renormalized theory. If in the renormalized quantities the renormalization $Z^{-2}$ for the two propagators and the vertex renormalization $Z_4$ compensate each other, the RG transformation would not act on the couplings and the system would have therefore a line of fixed points in terms of their bare values, providing an anomalous behavior in $G$ only.

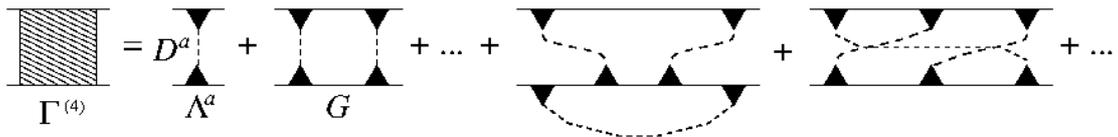

Fig. 3 - Perturbative corrections to the four point vertex $\Gamma^{(4)}$ (filled square): the vertex $\Lambda^a$ (black triangle) and the effective interaction $D^a$ (dashed line) always appear in the combination $(\Lambda^a)^2 D^a$.

The WI (9) allows us to eliminate the density vertex $\Lambda_0^a$ in favor of $G^{-1}$. We should therefore introduce $\Lambda_0^a$ in the argument instead of $\Gamma^{(4)}$. In analogy with the skeleton diagrammatic structure of quantum electrodynamics, where the role of the interaction is played by the photon propagator, any Green's or vertex function can be expressed in terms of the propagator $G$, the irreducible charge (spin) vertex $\Lambda_0^a$ and the effective interaction $D^a$, obtained as a resummation of the couplings $g^a$ (analogous to the photon propagator in quantum electrodynamics). In the skeleton structure of the four point vertex $\Gamma^{(4)}$, $\Lambda^a$ and $D^a$ appear in the combination $(\Lambda^a)^2 D^a$ (see Fig. 3). For small $q$, $D^a$ can be obtained as RPA summation of *bare* bubbles $\Pi_0^a$ for each coupling $g^a$, according to Fig. 4.

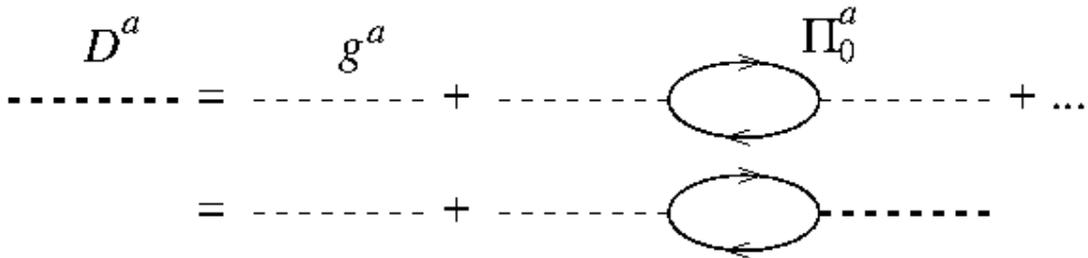

Fig. 4 - The effective interaction $D^a$ (thick dashed line) results from the RPA resummation of the bare interaction $g^a$ (thin dashed line) dressed by the bare fermion polarization bubble $\Pi_0^a$ (thin solid line).

It is enough to show that the renormalized polarization bubble $\Pi^a$ coincides with $\Pi_0^a$ for the non-interacting system. The vertex and selfenergy corrections of the renormalized bubbles indeed cancel each other due to Eq. (9), as depicted in Fig. 5.

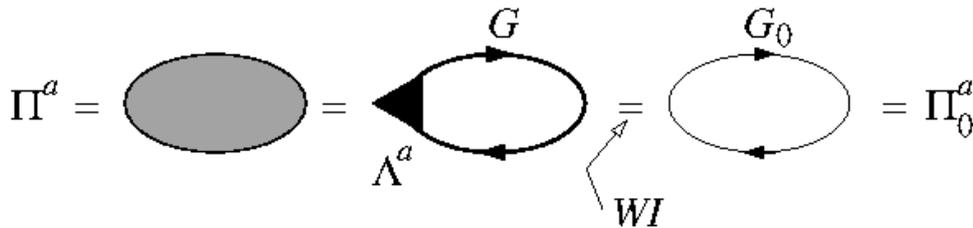

Fig. 5 - The cancellation of selfenergy and vertex corrections to the polarization bubble is enforced by the WI, Eq. (9). Here $G$ is the full propagator and $G_0$ is the free-particle propagator.

In conclusion, it is important to notice that the effective interaction $D^a$, resulting from the resummation in Fig. 4, has zero dimensionality, i. e., due to the WI (9), does not require to be renormalized and is marginal. On the other hand, the renormalization of the vertex $\Lambda^a$, again according to Eq. (9), is equal to the renormalization $Z$ of $G^{-1}$. Therefore, the four point vertex $\Gamma^{(4)}$ (see Fig. 3) renormalizes as $(\Lambda^a)^2$ i. e. $Z_4 = Z^2$, which therefore cancels the renormalization $Z^{-2}$ of the two propagators. This is the requirement to obtain a line of fixed points in terms of the bare couplings $g_i$. It can be further shown that no other renormalizations are required in the TL model [6,7]. This result assigns a different universality class to the Luttinger liquid with respect to the Fermi liquid.

## 2.5 Solution for the fermion Green's function.

Once the existence of the line of fixed points has been ascertained, a RG perturbative calculation of Z, and thus of the asymptotic behavior of G, could be obtained. For our purposes, it is however more convenient to proceed to calculating G directly from its Dyson equation with the help of the WI, Eq. (9) [7,10].

The effective interaction D, entering into the Dyson equation, resummed according to Fig. 4 and with respect to the index a, contains all the information of the charge and spin collective modes and all the couplings via their velocities $u^a = (v^a \tilde{v}^a)^{1/2}$ [7],

$$D(q) = \pi(\omega - \alpha v_F \mathbf{q}) \sum_{a=c,s} \left[ \frac{(2-\eta^a)(u^a - v_F)}{\omega - u^a \mathbf{q}} + \frac{\eta^a(u^a + v_F)}{\omega + u^a \mathbf{q}} \right], \quad (10)$$

where $\eta^a = [(v^a/\tilde{v}^a)^{1/2} + (\tilde{v}^a/v^a)^{1/2} - 2]/4$ and $\eta = \eta^c + \eta^s$, as it is shown in the next section, turns out to be the anomalous index of G, which depends on the bare couplings through the velocities.

Once D is known, since the density vertex can be eliminated by means of the WI, Eq. (9), the Dyson equation for the fermion propagator (Fig. 6) becomes a closed integral equation

$$(\omega - \alpha v_F \mathbf{p}) G(p) = 1 - \int \frac{d\mathbf{q} d\omega'}{(2\pi)^2} \frac{D(q) G(p-q)}{\omega - \omega' - \alpha v_F \mathbf{q}}. \quad (11)$$

The solution of Eq. (11) can be obtained after a Fourier transform to real space and time, and has the form

$$G(\mathbf{r},t) = e^{L(\mathbf{r},t) - L(0,0)} G_0(\mathbf{r},t), \quad (12)$$

where $L(\mathbf{r},t)$ is the Fourier transform of $iD(q)[\omega - \alpha v_F \mathbf{q} + i0^+ \text{sgn}(\omega)]^{-2}$, and the free-particle Green's function is $G_0(\mathbf{r},t) = (1/2\pi)[|\mathbf{r}| - v_F t + i0^+ \text{sgn}(t)]^{-1}$. Using Eq. (10) for D, it is found that $L(\mathbf{r},t)$ behaves logarithmically in $|\mathbf{r}|$ and t. Since $L(\mathbf{r},t)$ appears in the exponent of Eq. (12), G behaves as a power law, with an anomalous exponent that depends on $g_2$ and $g_4$ through $v^a$ and $\tilde{v}^a$. The single-particle excitation density and the discontinuity in the momentum distribution of the quasiparticles vanish at the Fermi surface with the anomalous exponent $\eta$.

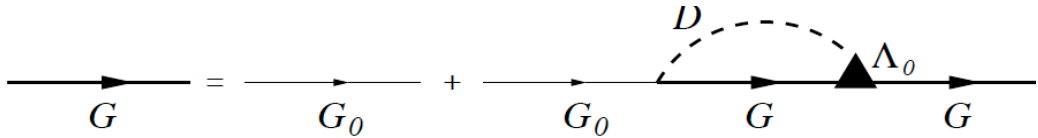

Fig. 6 – Dyson equation for the fermion Green's function.

The single-particle propagation is realized in a complex way, and results from the superposition of collective charge and spin modes, which propagate with different velocities (charge and spin separation).

# 3. Crossover of the Luttinger liquid to the Fermi and anomalous liquids.

## 3.1 Crossover with short-range interactions.

The effect of increasing the space dimensionality on the Luttinger liquid can be evaluated by extending the WI approach to continuous dimension $1 \leq d < 2$ [7,12]. As generalization of the TL model in $d > 1$, we assume a model Hamiltonian for the low-lying excitations close to the Fermi surface with dominating forward small $q$ scattering processes ($|\mathbf{q}| < \Lambda \ll k_F$).

The extension of the theory to non-integer dimension $1 \leq d < 2$ is accomplished, as customary, by analytical continuation of the Feynman diagrams to the complex $d$ plane. It is sufficient to continue momentum integrals of functions $f$ that depend on the momentum $\mathbf{k}$ only via $|\mathbf{k}|$ and the angle $\theta$ between $\mathbf{k}$ and another fixed momentum $\mathbf{p}$. Then

$$\int d^d \mathbf{k}\, f(\mathbf{k}) = S_{d-1} \int d|\mathbf{k}|\, |\mathbf{k}|^{d-1} \int_0^\pi d\theta (\sin\theta)^{d-2} f(|\mathbf{k}|, |\mathbf{k}||\mathbf{p}|\cos\theta), \qquad (13)$$

where $S_d$ is the surface of the unit sphere in $d$ dimensions. When $d \to 1$, $S_{d-1} \approx d-1$, and $S_{d-1}(\sin\theta)^{d-2} \to \delta(\theta) + \delta(\pi - \theta)$. The steps leading in $d=1$ to the additional WI's, Eqs. (8) or (9), are no longer strictly valid for $d>1$. Nonetheless, for $d<2$, both equations are still asymptotically valid, when the exchanged momenta $\mathbf{q}$ are small, since the typical integrals in Eq. (13) are peaked at $\theta=0, \pi$. In particular, since the relevant $\mathbf{k}$ vectors are asymptotically bound to be either parallel or antiparallel, near the Fermi surface we can still express the current vertex in terms of the density vertex, like in Eq. (8),

$$\vec{\Lambda}(p,q) = v_F \hat{\mathbf{p}} \Lambda_0(p,q), \qquad (14)$$

where $\vec{\Lambda}$ and $\Lambda_0$ are the current and density vertex, respectively, $\hat{\mathbf{p}} = \mathbf{p}/p_F$, and

$$\Lambda_0(p,q) = \frac{G^{-1}(p+q/2) - G^{-1}(p-q/2)}{\omega - v_F \hat{\mathbf{p}} \cdot \mathbf{q}}. \qquad (15)$$

Inserting Eq. (15) into the Dyson equation, one obtains again the Green's function in a form similar to Eq. (12), where however $L(\mathbf{r},t)$ is now the Fourier transform of $iD(q)[\omega - v_F q_r + i0^+ \text{sgn}(\omega)]^{-2}$, and $q_r = \mathbf{q} \cdot \hat{\mathbf{p}}$ is the *radial* component of $\mathbf{q}$. The expression for $L(\mathbf{r},t)$ involves now an average of the effective interaction $D(q)$ over the $d-1$ components of the transverse momentum,

$$\bar{D}_\Lambda(q_r,\omega) = \frac{S_{d-1}}{(2\pi)^{d-1}} \int_0^{\sqrt{\Lambda^2 - q_r^2}} dq_t\, q_t^{d-2} D\left(\frac{q_r}{\omega}, \frac{q_t}{\omega}\right). \qquad (16)$$

When $d \to 1$ one recovers the exact expression for $D$ in the TL. When $1 < d < 2$, the effective interaction, which was marginal for $d=1$, scales to zero in the infrared, since Eq. (16) implies the scaling relation $\bar{D}(q_r,\omega) \approx \omega^{d-1} \bar{D}(q_r/\omega)$. This result clearly illustrates the marginality of scattering processes with small transferred momenta in $d=1$, and their irrelevance in higher dimensions, in the case of regular (i.e., short ranged) scattering, as in the normal Fermi liquid.

## 3.2 Singular interactions.

Singular interactions can compensate the vanishing of the effective interaction $\bar{D}$ as $\omega$ goes to zero, and extend in this way the non-Fermi-liquid behavior to dimension grater than one.

According to our previous discussion, in $1<d<2$ forward processes still dominate the scattering at low energy, and the additional WI is asymptotically valid. In systems with short-range e-e interaction, however, when the e-e interaction is dressed by the collective modes, the integral over the transverse momentum leads to a suppression of the mixing between the single particle and the collective modes. The system crosses over to a FL behavior.

A non-FL behavior can be achieved in the presence of a sufficiently singular long-range (LR) e-e interaction [13,14,16], which we write in the form

$$V_{LR} = \frac{(g_{LR})^2}{2}\frac{1}{q^\alpha}.$$

This interaction strongly depends on the exchanged wavevector $q$ (as a power law with exponent $-\alpha$); its strength is given in terms of the coupling $g_{LR}$. As we have seen in Sec. 2.4, the bare effective interaction $V_{LR}$ (playing here the role of the $g$'s in the $g$-ology model), must be dressed by the RPA series. Again, as in the $d=1$ case, the irrelevance of selfenergy and vertex corrections to the RPA resummation in the presence of singular interaction can be confirmed by means of a RG analysis [14]. The RPA resummation with the bare bubbles acquires a complicated momentum and frequency dependence. For energies $\omega < v_F q$, the effective interaction becomes short-ranged, due to the particle-hole screening. However, in the opposite limit, $\omega > v_F q$, the effective interaction becomes singular,

$$D_{LR}(\mathbf{q},\omega) = \frac{1}{q^\alpha}\frac{\omega^2}{\omega^2 - c^2 q^{2-\alpha}}, \qquad (17)$$

with $c \propto g_{LR}$. The poles of $D_{LR}$ describe undamped collective plasmon excitations, which stay gapless for arbitrary $d$, provided $\alpha<2$, as it will be assumed henceforth. The dimension of momenta remains $[q]=1$, whereas the dimension of frequency is now dictated by the pole of the Eq. (17), and is $[\omega]\equiv z=1-\alpha/2$. We notice that the bare fermion propagator has dimension $-z$ instead of $-1$ in the dominating regime, where $D_{LR}$ has dimension $-\alpha$. The power counting of the four-point vertex correction in Fig. 1(b) gives now $d+z-a-2z\equiv d-d_C$, and the marginality is shifted to $d_C\equiv 1+\alpha/2>1$. Indeed, if we substitute $D$ with $D_{LR}$ in the integral Eq. (11) for the one particle Green's function, continued to $d$ dimensions, $L(\mathbf{r},t)$ appearing in its solution (12) behaves logarithmically (as in the Luttinger liquid for $d=1$) when $d=d_C$. As in the Luttinger liquid, thanks to the WI (9), $Z^2$ and the four-point vertex renormalization $Z_4$ compensate each other and the effective fermion-fermion coupling $u \propto (g_{LR})^2$ scales with its bare dimension $d_C-d$. An even stronger violation of the Fermi-liquid picture is found when $d<d_C$, where $L(\mathbf{r},t)$ has a power law singularity and the wavefunction renormalization $Z$ vanishes exponentially while $u$ scales to strong coupling. For $d>d_C$, $u$ vanishes in the infrared and the Fermi liquid is recovered.

## 4. Avoided quantum criticality in cuprates.

As we have seen in the previous section, a normal Fermi liquid is obtained as soon as $d>1$ for generic interacting Fermi particles, unless a singular potential compensates the reduction of mixing of the single particle with the collective modes. This is the case when the interaction is mediated by critical collective modes near instability. This is one of the reasons that led to look in the cuprates for a QCP separating two states underlying superconductivity. This QCP would indeed occur as a transition point at $T=0$, between two competing ground states, if superconductivity would not take over and avoid this quantum criticality (see Fig. 7).

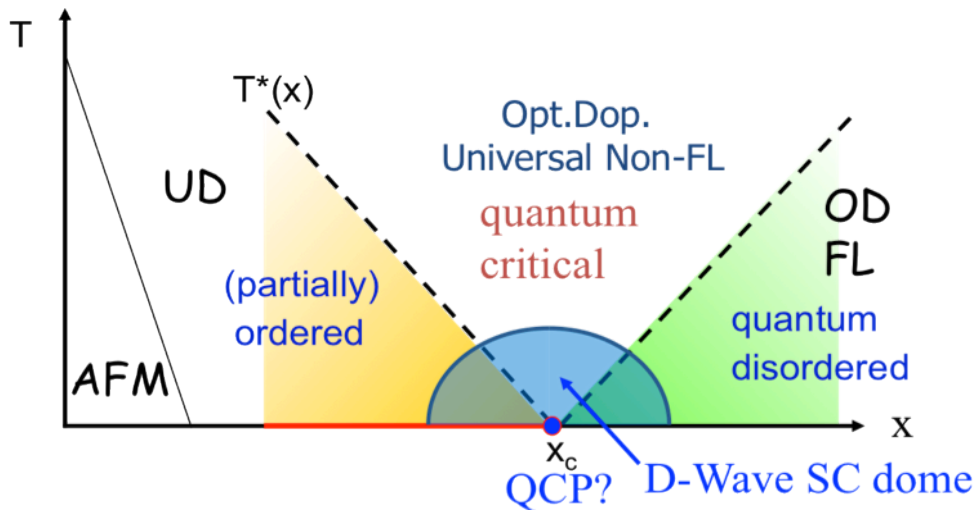

Fig. 7 – Schematic phase diagram of high-$T_c$ superconducting cuprates in the temperature $T$ vs. doping $x$ plane. The stoichiometric ($x=0$) and slightly doped cuprates are antiferromagnetic (AFM) insulators. An avoided QCP at $x=x_c$, hidden underneath thee superconducting dome, naturally divides the phase diagram into three regions: the underdoped (UD) (partially) ordered region, for $T<T^*(x)$; the quantum critical region, around optimal doping (Opt. Dop.), where the only energy scale is the temperature; the overdoped (OD) quantum disordered region, where the system behaves as a Fermi liquid in the metallic phase.

In quantum criticality [27] not only the symmetry of the local order parameter enters, but also its dynamics and therefore the collective modes are inextricably mixed in the generalized Landau-Wilson functional. In general, near a quantum phase transition, abundant dynamical long-range and long-time fluctuations exists which may naturally dress the fermionic quasiparticles, mediating an effective retarded e-e interactions which strongly depends on momentum, as well as on the control parameters (temperature and doping) ruling the proximity to criticality.

The most studied example of avoided QCP and non-Fermi-liquid behavior in the normal phase is the case of heavy fermion metals like $CePd_2Si_2$ (see, e.g., Ref. [28]). In this case, the competing states are a Kondo-compensated Fermi-liquid state and an antiferromagnetic state. Spin-density-wave fluctuations, the relevant collective modes, dress the fermionic quasiparticles leading to non-Fermi-liquid behavior and also act as the glue for the onset of superconductivity. Electrons reorganize themselves to give the superconducting state and avoid the QCP.

A similar situation has been suggested to occur in cuprates by several experimental results, both in thermodynamic and transport measurements (see, e.g., Ref. [29]). In particular, the crossover line $T^*(x)$, mentioned in Sec. 1.3, which separates the anomalous metallic phase from the pseudogap phase, extrapolates to $T^*=0$ K at around optimal doping ($x \approx 0.15 - 0.20$). However, the nature of the competing order and of the related critical mediators is still debated [21,22,30—34].

Unraveling the dynamics of specific quantum criticality, and the eventual competitions leading to it, is one hot research lines and the starting point for future, more sophisticated, field theoretical approaches. In particular, in cuprates, disentangling the relevant modes of the underlying state on which high-temperature superconductivity establishes, is a relevant current issue, as well as how these modes, coupling to fermionic quasiparticles, may account for the violation of the Fermi liquid and act as the glue for pairing.

Along the years, in our group, we elaborated on the idea that strongly correlated electron systems are on the verge of an instability towards phase separation in high- and low-density regions. Strong correlation, forbidding double occupancy of two fermions on the same site, reduces the homogenizing kinetic energy contribution. This, in turn, enhances the charge susceptibility and favors phase separation. In the absence of long-range Coulomb interactions, a wealth of mechanisms, which are ineffective in ordinary metals (e.g., short-range exchange, or coupling to phonons), can make the system unstable. Indeed, phase separation is observed in cuprates, whenever mobile ions are present to compensate for the electron charge unbalance. In phase separation, near criticality, the effective interaction is singular and stronger than Coulomb, and spoils the Fermi-liquid behavior [14]. Its occurrence should establish the onset of the heterogeneous phase below $T^*(x)$. However, in most cases, long-range Coulomb interactions frustrate macroscopic phase separation. As a compromise, mesoscopic charge segregation or an incommensurate charge density wave (CDW) may occur in the system. This has indeed been found in the Hubbard-Holstein model with long-range Coulomb interaction [21]. The modulation of the CDW was characterized by a finite wavevector $\mathbf{Q}_c \approx (\pm\pi/2, 0); (0, \pm\pi/2)$, and the avoided QCP was located near optimal doping, $x_{QCP} \approx 0.19$. According to our proposal, the avoided incommensurate charge-density-wave instability line $T_{CO}(x)$ should be identified with the line $T^*(x)$ [36].

Upon underdoping, the CDW should gradually evolve into the highly anharmonic charge profile of the stripe phase observed in some cases by neutron scattering [23]. Moreover, we recall that the metallic phase of cuprates is obtained from the antiferromagnetic undoped parent compound by chemical doping. Persistent incommensurate spin fluctuations with characteristic wavevector close to the ordering wavevector of the antiferromagnetic phase $\mathbf{Q}_s \approx (\pi,\pi)$ are expected to be enhanced within the charge-poor domains promoted by CDW formation. The region of existence of nearly critical spin fluctuations is in this way extended even in the optimally and slightly over-doped regions of the phase diagram, far from the antiferromagnetic QCP, located at low doping ($x_{AFM} \approx 0.05$). A continuous evolution upon doping of the relative importance of charge and spin fluctuations is then present, the former dominating around optimal doping, the latter dominating in the underdoped region.

We will not recall here the complete scenario of high-$T_c$ superconductors stemming from the proposal of quantum criticality related to the onset of a heterogeneous phase. According to the line followed in this work, we only address the question if we can detect the relevant modes, which would be implied in building up the effective singular interaction. This was recently achieved [24] by suitably interpreting the experiments of Raman spectroscopy in the $La_{2-x}Sr_xCuO_4$ family of cuprates. Strongly interconnected nearly critical charge and spin fluctuations are present, as implied by the above picture of incommensurate charge-density-wave quantum criticality.

To be unbiased, we assume that both charge and spin collective mode propagators take the form of a dynamical Ornstein-Zernicke propagator within the Gaussian approximation,

$$D_a(\mathbf{q},\omega_n) = \frac{1}{m_a + v_a(\mathbf{q}-\mathbf{Q}_a)^2 + |\omega_n| + \frac{\omega_n^2}{\overline{\Omega}_a}}, \qquad (18)$$

where henceforth the label $a=c,s$. The mass $m_a$ is proportional to the inverse square correlation length for charge (spin) fluctuations, $v_a$ is a scale controlling the dispersion of the collective mode, the scale $\overline{\Omega}_a$ separates the low-frequency Landau-damped diffusive regime from the high-frequency propagating regime and $\omega_n$ is the bosonic Matsubara frequency. Notice that the effective interaction $D$ in the extension of Luttinger liquid to singular interaction (see Sec. 3.2) was gapped when $\alpha \geq 2$. Here, the role of the gap is played by the mass $m_a$, which vanishes at criticality and the effective interaction is sufficiently singular to destroy the Fermi liquid, at least in some regions of the Fermi surface [37]. Indeed, when the fermion quasiparticles are coupled to critical collective modes with finite characteristic wavevectors $\mathbf{Q}_a$, the renormalization $Z$ of the single particle Green's function vanishes at the so-called hot spots, i.e., the points of the Fermi surface connected by $\mathbf{Q}_a$ (see Fig. 8). The resulting violation of the Fermi liquid cannot however be inferred directly from the scheme of Sec. 3.2, valid for forward singular interaction, since the modulating $\mathbf{Q}_a$ vectors introduce dominant scattering with large momentum transfer. The question whether a violation of the Fermi-liquid behavior at some points of the Fermi surface only would be enough to account for the non-Fermi-liquid properties of cuprates has been questioned [38]. Indeed, one could argue that transport properties are dominated by quasiparticles with momenta away from the hot spots, described by a Fermi-liquid propagator with finite $Z$. However, in the present theory with both charge and spin quasi-critical collective modes, the Fermi surface is disseminated with hot spots (24 hot spots due to spin and 16 hot spots due to charge collective modes are found for typical values of $\mathbf{Q}_s$ and $\mathbf{Q}_c$, and for the Fermi surface suited to $La_{2-x}Sr_xCuO_4$ systems [24]). Then, the argument about the dominance of Fermi-liquid quasiparticles is by no means conclusive. As we discuss below, the present theory, although with a completely different microscopic realization, yields a violation of the Fermi liquid that closely resembles the marginal-Fermi-liquid scenario [39]. This scenario was phenomenologically introduced to describe a non-Fermi-liquid quasiparticle inverse lifetime linear in energy and temperature.

In Ref. [24] the crucial observation was made that the effect of collective modes with finite characteristic wavevector on the fermionic quasiparticles could be investigated and resolved by means of Raman spectroscopy. On the one hand, Raman spectroscopy probes different regions of the Fermi surface by selecting the polarization of the ingoing and outgoing photons. On the other hand, due to the specific value of the characteristic wavevctor $\mathbf{Q}_a$ connecting different regions of the Fermi surface, specific asymptotic cancellations occur in the Raman response, depending on the relative sign of the Raman vertices involved (see Fig. 8). From this dependence it will follow that, at low frequency, the charge collective mode affects the Raman response in the so-called $B_{2g}$ channel [with Raman vertex $\gamma_k=\sin(k_x)\sin(k_y)$] whereas the spin collective mode contributes to the $B_{1g}$ channel [with Raman vertex $\gamma_k=\cos(k_x)-\cos(k_y)$].

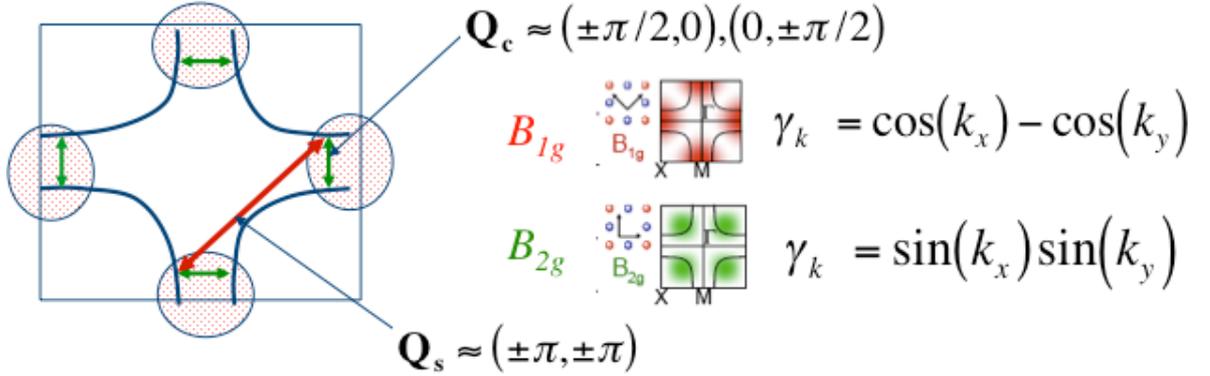

Fig. 8 – Sketch of the Fermi surface typical of the $La_{2-x}Sr_xCuO_4$ cuprate family, with examples of hot spots associated to the characteristic wavevectors of charge ($\mathbf{Q}_c$) and spin ($\mathbf{Q}_s$) fluctuations. The structure of the Raman vertices $\gamma_k$ in the $B_{1g}$ and $B_{2g}$ channels is also shown in the small insets: the red (green) areas are the regions where the vertex in the $B_{1g}$ ($B_{2g}$) channel is large in absolute value and the white areas are the region where the vertex changes sign. The wavevector $\mathbf{Q}_c$ connects region of the Fermi surface where $\gamma_k$ has the same sign in the $B_{1g}$ channel and changes sign in the $B_{2g}$ channel. The opposite happens for the wavevector $\mathbf{Q}_s$.

Indeed, within a standard memory function approach [40], the Raman response function can be written as

$$\chi(\omega) = \frac{\chi_0 \omega}{\omega + M(\omega)},$$

where $\chi_0$ is a real constant and $M(\omega)$ is the complex memory function resulting from scattering processes mediated by the propagator $D_a$, Eq. (18). Within a perturbative approach, the first contributions to $M(\omega)$ come from selfenergy and vertex corrections associated with the exchange of one collective mode, dressing the free-electron Raman bubble (see Fig. 9).

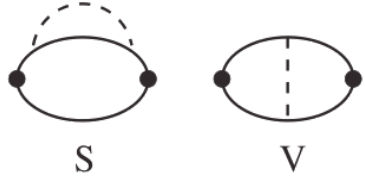

Fig. 9 - Selfenergy (S) and vertex (V) corrections to the Raman response, due to the exchange of a collective mode propagator, Eq. (18), (dashed line). The solid line represents the bare fermion propagator and the black circle represents the Raman vertex $\gamma_k$.

At this lowest order in perturbation theory, the charge and spin collective modes do not mix and contribute independently, i.e., $M(\omega)=M_c(\omega)+M_s(\omega)$. Assuming that close to criticality the perturbative corrections are dominated by the pole of the collective mode propagator, Eq. (18), one finds

$$\mathrm{Im}\,M_a(\omega) = \frac{1}{\omega}\int_0^\infty dz\,[\alpha^2 F(z)] \sum_{p=\pm 1}\left[\omega\coth\left(\frac{z}{2T}\right) - p(z+p\omega)\coth\left(\frac{z+p\omega}{2T}\right)\right],$$

which has the standard form of an inverse lifetime due to the scattering with a bosonic collective mode, and $\alpha^2 F_a$ is the spectral function of the charge ($a=c$) and spin ($a=s$) collective modes. In the present case, direct calculation of the diagrams in Fig. 9 gives

$$\alpha^2 F_a(\omega) = g_a\left[\arctan\left(\frac{\overline{E}_a}{\omega}-\frac{\omega}{\overline{\Omega}_a}\right) - \arctan\left(\frac{m_a}{\omega}-\frac{\omega}{\overline{\Omega}_a}\right)\right], \qquad (19)$$

where $\overline{E}_a$ is the energy cutoff of the collective mode dispersion and the dimensionless coupling

$$g_a \propto \gamma_{HS}(\gamma_{HS}-\gamma_{HS'})$$

depends on the Raman vertices calculated at two hot spots connected by the characteristic wavevector $\mathbf{Q}_a$. This expression enforces the cancellation quoted above: due to the specific values of $\mathbf{Q}_s$ and $\mathbf{Q}_c$, $\gamma_{HS}-\gamma_{HS'}$ vanishes in the $B_{1g}$ channel for the charge collective mode and in the $B_{2g}$ channel for the spin collective mode (see Fig. 8), at least at low frequency, where the dominant pole approximation is asymptotically valid. Therefore, the analysis of the Raman response in the two channels allows disentangling the contribution of the two collective modes, which would instead contribute jointly, e.g., in optical and ARPES spectra.

Eq. (19), resulting from coupling the fermionic quasiparticles with the charge and spin modes, provides a generalization of the so-called gapped marginal-Fermi-liquid spectral function [41], and formally reduces to it in the limit $\overline{\Omega}_a \to 0$ (with $\overline{E}_a\overline{\Omega}_a$ and $m_a\overline{\Omega}_a$ finite). In this limit, the spectral function assumes a box-like shape, and this implies that the fermionic quasiparticle inverse lifetime increases linearly with frequency as long as the frequency is inside the box and saturates to a constant outside the box. The uncertainty of the quasiparticle is equal to its energy value, showing a "marginal" violation of Fermi liquid behavior.

By a systematic fit of the Raman data for various temperatures (50 K<$T$<200 K) and doping levels (0.15<$x$<0.26), the parameters of the two collective modes and their evolution were extracted. The overall spectral weight $W_a$ of each collective mode is obtained integrating Eq. (19) over all frequencies. $W_c$ increases with increasing doping and tends to saturate in the most overdoped samples, whereas $W_s$ is larger in the less doped sample and decreases with increasing doping (see Fig. 10).

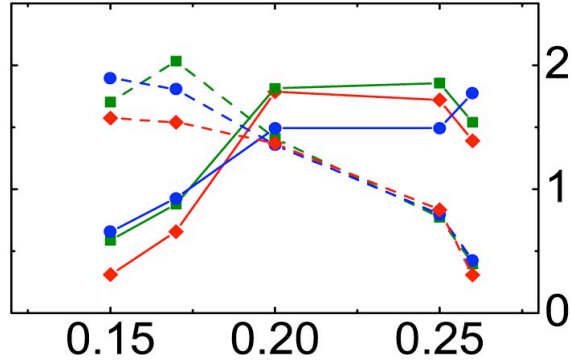

Fig. 10 - Spin (dashed lines) and charge (full lines) spectral weight $W_s$ and $W_c$ (in units of $10^3$ cm$^{-1}$) as a function of doping $x$, for $T$=50 K (blue line, circles), $T$=100 K (green line, squares), and $T$=200 K (red line, diamonds).

The spectral intensity of the two modes is equal at doping $x \approx 0.19$. Thus, the fermionic quasiparticles are more strongly scattered by spin collective modes for $x <$ 0.19 and by charge collective modes for $x >$ 0.19. At the doping $x \approx 0.19$, separating the two regions, we also obtained a mass of the charge collective mode linearly vanishing in temperature, making the temperature the only relevant energy scale, as it would implied by the presence a QCP at this doping.

Thus, our analysis supports the proposal that the phase competing with superconductivity in the underdoped region is characterized by (nearly) ordered stripe-like textures, whose onset on the overdoped side is marked by an incommensurate CDW instability. The related charge and spin fluctuations mediate a singular retarded effective e-e interaction, which accounts for anomalous physical responses. As in the case of heavy fermions, at low temperature the charge carriers reorganize themselves into a superconducting phase and avoid the occurrence of the QCP for charge ordering.